\documentclass[conference]{IEEEtran}
\IEEEoverridecommandlockouts
\usepackage{cite}
\usepackage{amsmath,amssymb,amsfonts}
\usepackage{algorithmic}
\usepackage{graphicx}
\usepackage{textcomp}
\usepackage{xcolor}
\usepackage{siunitx}
\DeclareSIUnit{\belmilliwatt}{Bm}
\DeclareSIUnit{\dBm}{\deci\belmilliwatt}
\usepackage{tabularx}
\newcolumntype{b}{>{\hsize=0.7\columnwidth}X}
\newcolumntype{w}{>{\hsize=0.2\columnwidth}X}
\ifCLASSOPTIONcompsoc
  \usepackage[caption=false,font=normalsize,labelfont=sf,textfont=sf]{subfig}
\else
  \usepackage[caption=false,font=footnotesize]{subfig}
\fi
\def\BibTeX{{\rm B\kern-.05em{\sc i\kern-.025em b}\kern-.08em
    T\kern-.1667em\lower.7ex\hbox{E}\kern-.125emX}}
\begin{document}

\title{Evaluation of HTTP/DASH Adaptation Algorithms on Vehicular Networks\\
}

% This condensed style is taken by a different conference template
%	\author{
%		\IEEEauthorblockN{Dimitrios~J.~Vergados\IEEEauthorrefmark{1}, Angelos~Michalas\IEEEauthorrefmark{1}, Katina Kralevska\IEEEauthorrefmark{2}, 
%		          and Dimitrios~D.~Vergados\IEEEauthorrefmark{3}\\ Email: djvergad@gmail.com, amichalas@kastoria.teikoz.gr, katinak@ntnu.no, vergados@unipi.gr}
%		\IEEEauthorblockA{\IEEEauthorrefmark{1}Department of Informatics Engineering, Technological Education Institute of Western Macedonia, Kastoria, Greece}
%		\IEEEauthorblockA{\IEEEauthorrefmark{2}Department of Information Security and Communication Technology, 
%			NTNU, Norwegian University of Science and Technology}
%		\IEEEauthorblockA{\IEEEauthorrefmark{3}Department of Informatics, University of Piraeus, Piraeus, Greece}
%	}

\author{\IEEEauthorblockN{Dimitrios J. Vergados}
\IEEEauthorblockA{\textit{Dept. of Informatics Engineering} \\
\textit{Western Macedonia University of Applied Sciences}\\
Kastoria, Greece\\
djvergad@gmail.com}\\
\IEEEauthorblockN{Katina Kralevska}
\IEEEauthorblockA{\textit{Dept.of Information Security and Communication Technology} \\
\textit{Norwegian University of Science and Technology}\\
Trondheim, Norway \\
katinak@ntnu.no}
\and
\IEEEauthorblockN{Angelos Michalas}
\IEEEauthorblockA{\textit{Dept. of Informatics Engineering} \\
\textit{Western Macedonia University of Applied Sciences}\\
Kastoria, Greece\\
amichalas@kastoria.teiwm.gr}\\
\IEEEauthorblockN{Dimitrios D. Vergados}
\IEEEauthorblockA{\textit{Dept. of Informatics} \\
\textit{University of Piraeus}\\
Piraeus, Greece \\
vergados@unipi.gr}
}

\maketitle

\begin{abstract}
Video streaming currently accounts for the majority of Internet traffic.
One factor that enables video streaming is HTTP Adaptive Streaming (HAS),
that allows the users to stream video using a bit rate that closely matches
the available bandwidth from the server to the client. 
MPEG Dynamic Adaptive Streaming over HTTP (DASH) is a widely used standard,
that allows the clients to select the resolution to download based on their
own estimations. 
The algorithm for determining the next segment in a DASH stream is not part
of the standard, but it is an important factor in the resulting playback quality.
Nowadays vehicles are increasingly equipped with mobile communication
devices, and in-vehicle multimedia entertainment systems.
In this paper, we evaluate the performance of various DASH adaptation algorithms
over a vehicular network.
We present detailed simulation results highlighting the advantages and
disadvantages of various adaptation algorithms in delivering video content
to vehicular users, and we show how the different adaptation algorithms perform
in terms of throughput, playback interruption time, and number of interruptions.

\end{abstract}


\section{Introduction}
Fifth generation (5G) Vehicular Ad hoc Networks (VANETs) are expected to provide
modern services with different Quality of Service (QoS) characteristics.
To support the communication needs of vehicular users’, dense deployments of 5G
access networks are applied for the interaction between the vehicles and network
infrastructure such as 3GPP Long Term Evolution Advanced (LTE-A)~\cite{E-UTRAN}
macrocells and femtocells as well as IEEE 802.11p Wireless Access for Vehicular
Environment (WAVE)~\cite{Wave} Road Side Units (RSUs). 

Driver assistance services in VANETs usually demand the transmission of multimedia
data, occurring on a highway, including video clips of an accident or a critical
situation (e.g. traffic congestion, fire, flood, terrorist attack).
Moreover, passengers' entertainment and information services require the reception of
multimedia data in an acceptable quality.
However, the vehicular environment is particularly challenging for reliable streaming of multimedia content due to the following factors:
a)~the high speed of the users causes the channel conditions to change rapidly;
as they move into and out of areas of deep fading/low signal, and/or enter/leave
congested areas in rapid succession;
b)~the movement at high speeds causes Doppler effects that may hinder the wireless
transmission;
c)~as the speed of the vehicles increases, the handovers must be performed
quicker to prevent the interruption of the service.
To guarantee the reliability of services in vehicular environment, specific algorithms should be applied to adjust the video rate delivered to each user in accordance to the current
network conditions. 
We expect adaptation algorithms that work well in wired/static wireless
settings to have problems in maintaining both high video quality and
uninterrupted streaming at the same time in a vehicular setting. 

In this paper, we focus on the MPEG Dynamic Adaptive Streaming over HTTP
(MPEG-DASH) (ISO/IEC 23009-1) standard, issued by MPEG in 2012, for HTTP
Adaptive Streaming (HAS) for provisioning of video services in VANETs.
MPEG-DASH allows users to access video streams of multiple resolutions
available at a server.
However, the standard does not define how the user could adapt to time
varying bandwidth in order to achieve better quality.

In our previous work~\cite{vergados2014, vergadospimrc2014, vergados2016fdash} we proposed a novel MPEG-DASH rate adaptation scheme, namely FDASH, aiming to efficiently adjust the video rate delivered to each user in accordance to the current network conditions. The clients automatically choose the bit rate representation of each video clip so that interruptions and frequent resolution changes are avoided. The current paper extends our previous work by comparing FDASH with other adaptation algorithms in the context of a vehicular network over LTE. In particular, we use the LENA scenario of ns3, that contains realistic settings for mobile users over an LTE network, with multiple access points, and both vehicular and fixed (residential) users, in order to compare the performance of the FDASH algorithm with other algorithms in the literature over an LTE vehicular scenario.
We evaluate the performance in terms of throughput, playback interruption time, and number of interruptions.
A thorough evaluation of the state-of-the-art HAS algorithms in 4G networks is crucial in order to follow up the pace of 5G and to meet the requirements of the upcoming vehicular applications.
The simulation results show that the FDASH algorithm shows an overall good performance in terms of all performance metrics compared to the other algorithms from the literature.

This paper is structured as follows. In Section~\ref{sec:related} we present a review of the relevant literature.
In Section~\ref{sec:fdash} the FDASH algorithm is presented.
Section~\ref{sec:evaluation} provides simulation results indicating the efficiency of the proposed model.
Finally, we conclude the paper and discuss ongoing and future research directions in Section~\ref{sec:conclusion}.

\section{Related Work}
\label{sec:related}

The problem of determining the optimal bit rate in an adaptive streaming has
been studied extensively in the literature. 
We focus on adaptation mechanisms that are performed entirely at the client 
side, using measured network conditions and the buffer occupancy.

The Adaptation Algorithm for Adaptive Streaming over HTTP (AAASH)~\cite{miller2012}
tries to optimize the user's experience by balancing the following goals:
a) to prevent video playback interruptions when possible;
b) to maintain a high average and minimum video resolution;
c) to decrease the number of resolution changes;
d) to minimize the initial buffering time when the playback starts.

%adapts the video rate requested by the client using a complex adaptation mechanism with multiple conditions and configuration parameters. 
%The algorithm downloads the first segment at the lowest representation to reduce the time until the playback starts.  
%Then, the video quality is increased by introducing a fast start phase at the beginning of the playback. It proceeds to the next phase in case one of the following conditions are met:
%a) the highest video resolution has been reached;
%b) the buffer level is not increased monotonically;
%c) the bit rate of the segment resolution is close to the measured throughout.

Reference~\cite{Liu2011} presents an adaptation scheme called Rate Adaptation
for Adaptive HTTP Streaming~(RAAHS).
RAAHS exploits the segment fetch time and compares it to the client's
playback time in order to calculate the bit rate of the following segment.
Switch up is done using a step-wise process, whereas switch down is done
in a single step (aggressive).
There is also a mechanism to limit the maximum buffering time.

%algorithm which evaluates the network bandwidth by comparing
%the segment fetch time and the playback time at the client.
%A step-wise switch up and an aggressive switch down method are used to adjust the download bit rate to the available throughput.
%Furthermore, an idle time is determined to control requests for video segments
%and confine the buffering time to a maximum limit.

The authors extended further the work by proposing the 
Serial and Parallel Segment Fetching Time Methods (SFTM/PFTM),
which is an 
algorithm for adapting the bit rate of DASH streams, taking
advantage of Content
Distribution Networks (CDNs)~\cite{Liu2012}.
In order to detect the network availability, they defined a rate adaptation metric as the ratio of the expected segment fetch time and the measured segment fetch time.
The expected segment fetch time takes into account the media segment duration and the buffering time at the client.
After detecting the network availability with the proposed rate adaptation metric, a step wise switch-up and a multi-step switch-down strategies are applied.
In addition, priority segment fetch times are assigned to new clients to improve fairness. 

The agile Smooth Video Adaptation Algorithm (SVAA) for DASH systems, proposed
in~\cite{Tian2012}, uses client-side buffered video time as feedback signal to estimate the video rate of the next downloadable video segment.
The algorithm increases smoothly the video rate with the available network bandwidth, and it reduces promptly the video rate in response to sudden congestion level shift-ups.
Moreover, it uses a rate margin to reduce slightly the video rate to limit video rate adjustments.
The buffer cap and the small video rate margin improve the smoothness in video rate and buffer size.

The authors in~\cite{Mok2012} replaced the original quality adaption algorithm in Adobe's Open Source Media Framework (OSMF) so that the quality level switching follows a pre-defined scenario.
The fetch times of last two video segments are used to estimate the 
bandwidth that is available between the server and the client.
This bandwidth estimation is then used to select the bit rate of the 
following segment, where the rate selected is the highest rate that 
is smaller than the estimated bandwidth.

In~\cite{8362846}, the authors applied a Markov chain to analyse the QoE metrics for the user,
namely the probability of the video to be interrupted, the initial buffering delay,
the average bit rate of the video, and the rate of bit rate changes.

The work in \cite{Ma17,e19090477} takes an advantage of a fuzzy logic application to handle the uncertainty of the network system. In \cite{e19090477}, the mobile QoS is improved by using a cumulative moving average in order to capture the related information between near-term past values and current values. %First, the average of previous bandwidth/buffer variation values of a client is determined. Second, a certain number of the previous fluctuation values to infer the degree of current network fluctuation are traced-back. Last, the next request representation for the client is predicted based on the averaged data.

In order to enhance user QoE in video distribution applications with DASH, mobile edge computing in LTE and 5G has been considered in \cite{Ge:2016:QDV:2984356.2988522,ge2017towards,8368984}. Caching enables storage of popular videos in the network edge, close to the users, however, caching cannot be applied to all videos.

Recent surveys \cite{7884970,8424813} give a good overview of the bit rate adaptation algorithms for DASH based content delivery.

\section{The FDASH Adaptation Algorithm}
\label{sec:fdash}

The idea behind the FDASH algorithm is to use a fuzzy controller to adapt
the video bit rate in a way that simultaneously prevents the video
playback from being disrupted due to buffer underflow and maintains
the highest video bit rate. 
The adaptation algorithm should quickly change the resolution, as the 
network conditions change due to mobility and changes in the amount of
background traffic.

To this end, our model is based on clients implementing the MPEG-DASH standard to request streams of video from an HTTP video server. A video stream residing at the server consists of $n$ segments of duration  $\tau$, and each segment is encoded in multiple resolutions of quality. Each client uses a fuzzy controller rate adaptation algorithm to estimate the resolution of the next video segment obtained from the server. The proposed algorithm tries to achieve the following: a) distribute the best possible resolution of video segments; b) deliver undisrupted video playback as a result of buffer underflows at the client; c) avoid unnecessary changes of video resolution owing to frequent fluctuations of the available connection throughput. 

The inputs that are used by the FDASH algorithm are the buffering time $t_i$,
and the change in buffering time  $\Delta t_i = t_i - t_{i-1}$ between
the last received segment and the previous one. The buffering time denotes the time $t_i$ that the last received segment $i$ waits at the client until it starts playing.

The output of the fuzzy controller ($f\left(t_i, 
\Delta t_{i-1}\right)$) is an increase/decrease factor,
which defines how much higher or lower the bit rate of the next
segment will be, compared to the estimated channel throughput over the last period.
Specifically,
\begin{equation}
b_k = f\left(t_i, \Delta t_{i-1}\right) \times r_d,
\end{equation}
where the term $r_{d}$ is the available connection throughput, estimated by taking into account the throughput $r_{i}, i=1 \ldots k,$ of the last $k$ segments downloaded during a specified period of time $d$ and taking the average. 
The definition of function $f$ encompasses the logic of the controller.
The one that is used for the FDASH algorithm is presented in~\cite{vergados2016fdash}.
Each segment throughput is estimated at the client as
\begin{equation}
r_i = \left(b_i \times \tau \right) / \left(t_i^e - t_i^b \right),
\end{equation}
where $b_i$ denotes the bit rate of segment $i$,  $t_i^e$ and $t_i^b$ denote the time when the segment $i$ has been started downloading and the time when the whole segment has been received at the client, respectively.

The final step of the algorithm tries to avoid unnecessary bit rate fluctuations. If $b_n > b_{k-1}$ and by selecting the new bit rate $b_k$, the buffer 
level is estimated
to be less than $T$ for the next $60$ sec, then the bit rate remains unchanged. 
Similarly, if $b_n < b_{k-1}$, but the old bit rate is estimated to 
produce a buffer level for the next $60$ sec that is larger than $T$, then 
the bit rate remains  unchanged. In all other cases, the bit rate of the next segment is set to $b_n$.

\section{Performance Evaluation}
\label{sec:evaluation}

\subsection{Simulation Scenario}

The simulation scenario that is evaluated in this paper consists of an LTE
network, following 3GPP R4-092042, Section 4.2.1 Dual Stripe Model, through the
use of the "lena-dual-stripe" ns3 example script. 
\footnote{The implementation of the DASH adaptation algorithm was performed using the ns3-dash
module we have developed and released publicly at https://github.com/djvergad/dash.
The simulation was run using ns3 version $3.28$.}

The scenario consists of a number of blocks of buildings, with several UEs moving at
vehicular speeds within the topology, while other UEs are located inside the buildings.
For each simulation run, each user downloads a video over LTE, where all users
use the same DASH adaptation algorithm, which is a parameter of the simulation.
The other simulation parameter is the maximum speed for the vehicular users.
We executed repeated iterations to produce average results, and we also calculated
the $95\%$ confidence interval for the mean.
The specific parameters of the scenario are listed in Table~\ref{table:parameters}.
We compare the following adaptation algorithms: FDASH, AAASH, OSMF, RAAHS, SFTM, and SVAA.

\begin{table}[htbp]
  \caption{Simulation Parameters}
  \begin{center}
    \begin{tabularx}{\columnwidth}{|b|w|}
      \hline
      \textbf{Parameter}&\textbf{Value} \\
%      \cline{2-4} 
      \hline
      Number of femtocell blocks                                 & 1    \\
      Number of apartments along the X axis in a femtocell block & 10   \\
      Number of floors                                           & 1    \\
      How many macro sites there are                             & 3    \\
      (Minimum) number of sites along the X-axis of the hex grid & 1    \\
      Min distance between two nearby macro cell sites           & \SI{500}{\meter} \\
      How much the UE area extends ouside the macrocell grid
      (expressed as fraction of the interSiteDistance)            & 0.5  \\
      How many macrocell UEs there are per square meter          & \SI{2e-5}{\meter^{-2}} \\
      The HeNB deployment ratio as per 3GPP R4-092042            & 0.2 \\
      the HeNB activation ratio as per 3GPP R4-092042            & 0.5 \\
      How many (on average) home UEs per HeNB there 
      are in the simulation                                      & 1.0 \\
      TX power used by macro eNBs                                & \SI{46.0}{\dBm} \\
      TX power used by HeNBs                                     & \SI{20.0}{\dBm} \\
      DL EARFCN used by HeNBs                                    & 100 \\
      Bandwidth [num RBs] used by macro eNBs                     & 25 \\
      Bandwidth [num RBs] used by HeNBs                          & 25 \\
      Total duration of the simulation                           & \SI{550}{\second} \\
      Resource Block Id of Data Channel, -1 means REM will be 
      averaged from all RBs of Control Channel                   & -1 \\
      epc: If true, will setup the EPC to simulate an end-to-end
      topology, with real IP applications over PDCP and RLC UM   & true \\
      epcDL: if true, will activate data flows in the downlink
      when EPC is being used                                     & true \\
      epcUL: If true, will activate data flows in the uplink when
      EPC is being used                                          & true \\
      useUDP: if true the UpdClient application will be used     & false \\
      useDash: if true the DashClient application will be used   & true \\
      fadingTrace                                                & false \\
      How many bearers per UE there are in the simulation        & 1 \\
      SRS Periodicity (has to be at least greater than the number
      of UEs per eNB)                                            & 80 \\
      Minimum speed value of macro UE with random waypoint model & \SI{1}{\meter\per\second} \\
      Maximum speed value of macro UE with random waypoint model & \SI{5\dots40 }{\meter\per\second} \\
      The target time difference between receiving and playing a
      frame                                                      & \SI{35}{\second} \\
      The window for measuring the average throughput            & \SI{10}{\second} \\
      \hline
%      \multicolumn{4}{l}{$^{\mathrm{a}}$Sample of a Table footnote.}
    \end{tabularx}
   \label{table:parameters}
  \end{center}
\end{table}

\subsection{Simulation Results}
\newcommand{\mywidth}{1.05}

\begin{figure}[htbp]
 \centerline{\includegraphics[width=\mywidth\columnwidth]{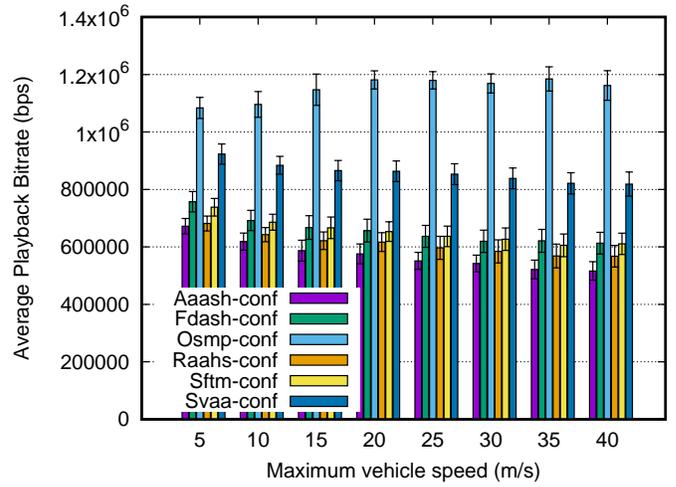}}
\caption{The average playback bit rate that is achieved for each algorithm and speed.}
\label{fig:lena_avg_rate}
\end{figure}

\begin{figure}[htbp]
\centerline{\includegraphics[width=\mywidth\columnwidth]{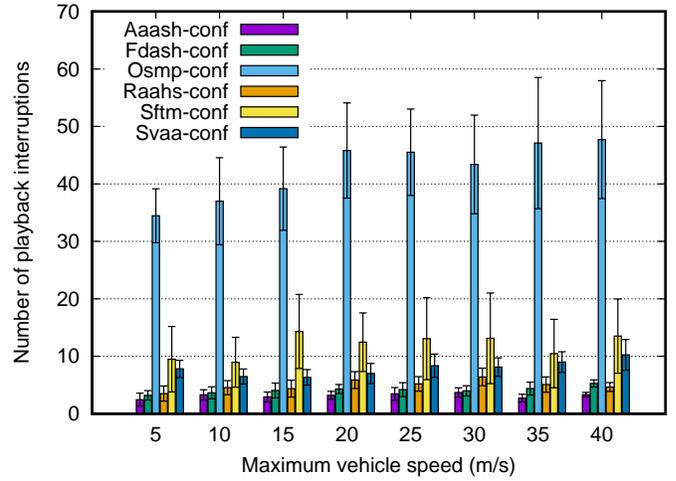}}
\caption{The average number of interruptions per video stream for each algorithm and speed.}
\label{fig:lena_interruptions}
\end{figure}

\begin{figure}[htbp]
 \centerline{\includegraphics[width=\mywidth\columnwidth]{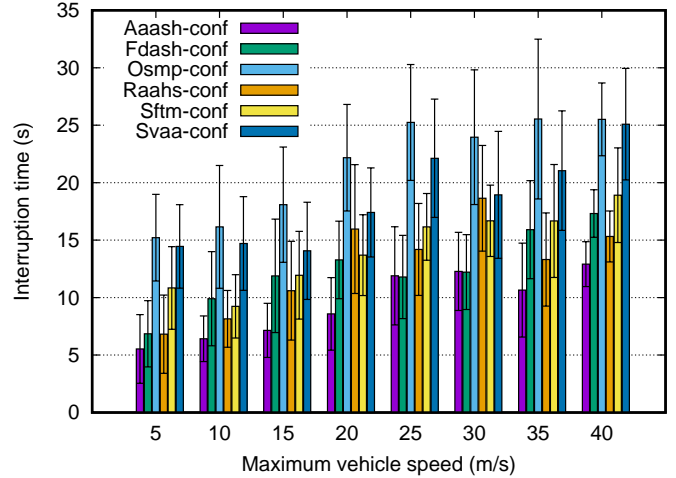}}
\caption{The average time of interrupted playback per video stream for each algorithm and speed.}
\label{fig:lena_interruption_time}
\end{figure}

\begin{figure}[htbp]
\centerline{\includegraphics[width=\mywidth\columnwidth]{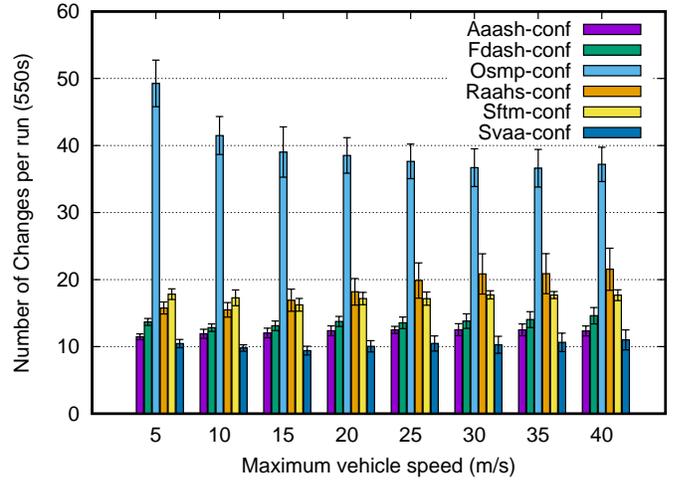}}
\caption{The average number of resolution changes per video stream for each algorithm and speed.}
\label{fig:lena_changes}
\end{figure}

\begin{figure}[htbp]
\centerline{\includegraphics[width=\mywidth\columnwidth]{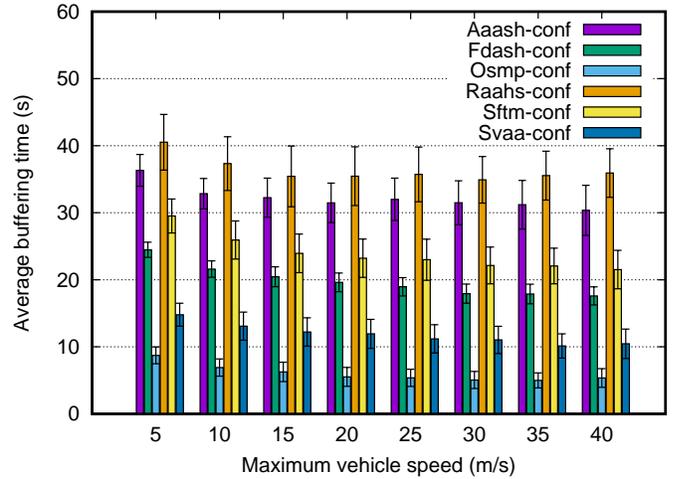}}
\caption{The average buffering time per video stream for each algorithm and speed.}
\label{fig:lena_avg_dt}
\end{figure}

The simulation results are depicted in Figs.~\ref{fig:lena_avg_rate}--\ref{fig:lena_avg_dt}.
Fig.~\ref{fig:lena_avg_rate} shows the average video playback bit rate that is
achieved for each algorithm. We can see that highest bit rate is achieved by OSMF,
then by SVAA and FDASH. Note however that the good performance in this 
metric by OSMF and SVAA comes as a result of worst performance in the other metrics.
The same figure shows that as the speed of the vehicles increases, the achieved rate reduces for all algorithms.

Specifically, from Fig.~\ref{fig:lena_interruptions}, OSMF has
the poorest performance in terms of number of interruptions compared to the other
algorithms.
FDASH and SVAA maintain good performance in terms of number of interruptions, despite
the high throughput that they achieve.

In terms of the time spent while the playback is interrupted, i.e. the interruption
time (Fig.~\ref{fig:lena_interruption_time}), we can see that the largest values
are obtained for SVAA and OSMF. It follows that these algorithms achieve large throughput,
but at a cost of high interruption time, so the users' QoE would be worse.
In our opinion, FDASH maintains the best balance between bit rate and interruption time.
From Fig.~\ref{fig:lena_interruption_time}, we can also observe that as the users increase their speed, the interruption time is also increased, as expected.

The next metric that we study is the number of resolution changes that are observed for each playback for the different algorithms (Fig.~\ref{fig:lena_changes}).
In general, users do not want to see the resolution to change too frequently,
but at the same
time, as network conditions change, some level of adaptation is unavoidable, unless the 
resolution is fixed at the lowest possible rate.
OSMF shows a larger number of resolution 
changes than the other algorithms by a large margin, with the lowest changes obtained
by SVAA.

Finally, we measure the average buffering time that each algorithm achieves 
(Fig.~\ref{fig:lena_avg_dt}).
For this metric there is also no optimal value, as lower buffering times make interruptions and/or resolution changes more likely. On the other hand, higher buffering times lead to lower playback bit rate, until the buffering time reaches its target. 
We can see that RAHHS and AAASH have the largest buffering time, with OSMF and SVAA
having the smallest, which is consistent with their relative performance in throughput
and interruptions. 
This relationship though is not always straight forward, for example FDASH has smaller
buffering time than both RAHHS and SFTM, but at the same time it has fewer interruptions
and a smaller or equal interruption times than these algorithms, especially when the 
speed of the UEs increases.

Overall, we can say that AAASH is the best in terms of the number of interruptions and the total interruption time, but it achieves the lowest bit rate and the 2nd largest buffering time. 
On the other hand, OSMP has the highest bit rate, but also has more interruptions and marginally the highest interruption time. Additionally, it has the highest number of resolution changes and the smallest buffering time.
SVAA achieves the second highest bit rate, the lowest resolution changes, and reasonable
number of interruptions, but the interruption time is the second largest. 
FDASH maintains a balance between high bit rates and low interruptions/interruption time,
while at the same the number of resolution changes are fairly limited and the buffering
time is low.

\section{Conclusion}
\label{sec:conclusion}
In this paper, we performed an evaluation of DASH adaptation algorithms over a vehicular
scenario, with respect to the effect of vehicle speed, on different quality of experience
metrics.
Although there are significant differences in the performance of the algorithms, 
we can say that none of the examined algorithms have a good enough performance for 
vehicular LTE networks, as in all cases, during a playback time of 550s we observed at least
3 interruptions, with the interruption time being more than 5 seconds on average for 
all the scenarios examined.
Thus, further research is needed to design adaptation algorithms that can perform flawlessly in a LTE vehicular setting.

Some directions for future extension of this work may include:
a)~to implement 
more adaptation algorithms in the simulation model;
b)~to perform a similar evaluation on a 5G simulation tool, as they become available;
c)~to explore how the device-to-device mode of 5G may help the video delivery, through the implementation of VANET;
d)~to explore if SDN can be used to distribute the resources among the vehicles in a way that there is both performance and fairness in terms of video quality. 
\bibliographystyle{IEEEtran}
 %\IEEEtriggeratref{2}
\bibliography{References}

\begin{thebibliography}{10}
\providecommand{\url}[1]{#1}
\csname url@samestyle\endcsname
\providecommand{\newblock}{\relax}
\providecommand{\bibinfo}[2]{#2}
\providecommand{\BIBentrySTDinterwordspacing}{\spaceskip=0pt\relax}
\providecommand{\BIBentryALTinterwordstretchfactor}{4}
\providecommand{\BIBentryALTinterwordspacing}{\spaceskip=\fontdimen2\font plus
\BIBentryALTinterwordstretchfactor\fontdimen3\font minus
  \fontdimen4\font\relax}
\providecommand{\BIBforeignlanguage}[2]{{%
\expandafter\ifx\csname l@#1\endcsname\relax
\typeout{** WARNING: IEEEtran.bst: No hyphenation pattern has been}%
\typeout{** loaded for the language `#1'. Using the pattern for}%
\typeout{** the default language instead.}%
\else
\language=\csname l@#1\endcsname
\fi
#2}}
\providecommand{\BIBdecl}{\relax}
\BIBdecl

\bibitem{E-UTRAN}
``{TS 36.213 version 14.2.0: Evolved Universal Terrestrial Radio Access Network
  (E-UTRAN) (Release 14)},'' 2017.

\bibitem{Wave}
``{1609.12-2016 - IEEE standard for wireless access in vehicular environments
  (wave) – networking services},'' 2016.

\bibitem{vergados2014}
D.~J. Vergados, A.~Michalas, A.~Sgora, and D.~D. Vergados, ``A control-based
  algorithm for rate adaption in mpeg-dash,'' in \emph{The 5th International
  Conference on Information, Intelligence, Systems and Applications}.\hskip 1em
  plus 0.5em minus 0.4em\relax IEEE, 2014, pp. 438--442.

\bibitem{vergadospimrc2014}
------, ``A fuzzy controller for rate adaptation in mpeg-dash clients,'' in
  \emph{The 25th IEEE International Symposium on Personal, Indoor and Mobile
  Radio Communications}.\hskip 1em plus 0.5em minus 0.4em\relax IEEE, 2014, pp.
  2008--2012.

\bibitem{vergados2016fdash}
D.~J. Vergados, A.~Michalas, A.~Sgora, D.~D. Vergados, and P.~Chatzimisios,
  ``Fdash: A fuzzy-based mpeg/dash adaptation algorithm,'' \emph{IEEE Systems
  Journal}, vol.~10, no.~2, pp. 859--868, 2016.

\bibitem{miller2012}
K.~Miller, E.~Quacchio, G.~Gennari, and A.~Wolisz, ``{Adaptation algorithm for
  adaptive streaming over HTTP},'' in \emph{19th International Packet Video
  Workshop}.\hskip 1em plus 0.5em minus 0.4em\relax IEEE, 2012, pp. 173--178.

\bibitem{Liu2011}
C.~Liu, I.~Bouazizi, and M.~Gabbouj, ``{Rate Adaptation for Adaptive HTTP
  Streaming},'' in \emph{Proceedings of the Second Annual ACM Conference on
  Multimedia Systems}, 2011, pp. 169--174.

\bibitem{Liu2012}
C.~Liu, I.~Bouazizi, M.~M. Hannuksela, and M.~Gabbouj, ``{Rate adaptation for
  dynamic adaptive streaming over HTTP in content distribution network },''
  \emph{Signal Processing: Image Communication}, vol.~27, no.~4, pp. 288 --
  311, 2012.

\bibitem{Tian2012}
G.~Tian and Y.~Liu, ``{Towards Agile and Smooth Video Adaptation in Dynamic
  HTTP Streaming},'' in \emph{Proceedings of the 8th International Conference
  on Emerging Networking Experiments and Technologies}, 2012, pp. 109--120.

\bibitem{Mok2012}
R.~K.~P. Mok, X.~Luo, E.~W.~W. Chan, and R.~K.~C. Chang, ``{QDASH: A QoE-aware
  DASH System},'' in \emph{Proceedings of the 3rd Multimedia Systems
  Conference}, 2012, pp. 11--22.

\bibitem{8362846}
S.~Poojary, R.~El-Azouzi, E.~Altman, A.~Sunny, I.~Triki, M.~Haddad, T.~Jimenez,
  S.~Valentin, and D.~Tsilimantos, ``Analysis of qoe for adaptive video
  streaming over wireless networks,'' in \emph{16th International Symposium on
  Modeling and Optimization in Mobile, Ad Hoc, and Wireless Networks (WiOpt)},
  May 2018, pp. 1--8.

\bibitem{Ma17}
L.~Ma, S.~Park, J.~Park, J.~Nam, J.~Jang, and J.~Kim, ``A fuzzy-based method
  for reducing mobile video-quality fluctuation,'' \emph{International Journal
  of Mobile Device Engineering}, vol.~1, pp. 21--28, 03 2017.

\bibitem{e19090477}
L.~V. Ma, J.~Park, J.~Nam, H.~Ryu, and J.~Kim, ``A fuzzy-based adaptive
  streaming algorithm for reducing entropy rate of dash bitrate fluctuation to
  improve mobile quality of service,'' \emph{Entropy}, vol.~19, no.~9, 2017.

\bibitem{Ge:2016:QDV:2984356.2988522}
C.~Ge, N.~Wang, S.~Skillman, G.~Foster, and Y.~Cao, ``Qoe-driven dash video
  caching and adaptation at 5g mobile edge,'' in \emph{Proceedings of the 3rd
  ACM Conference on Information-Centric Networking}, 2016, pp. 237--242.

\bibitem{ge2017towards}
C.~Ge, N.~Wang, G.~Foster, and M.~Wilson, ``Towards qoe-assured 4k
  video-on-demand delivery through mobile edge virtualization with adaptive
  prefetching,'' \emph{IEEE Transactions on Multimedia}, vol.~19, no.~10, pp.
  2222--2237, 2017.

\bibitem{8368984}
Y.~Tan, C.~Han, M.~Luo, X.~Zhou, and X.~Zhang, ``Radio network-aware edge
  caching for video delivery in mec-enabled cellular networks,'' in \emph{IEEE
  Wireless Communications and Networking Conference Workshops (WCNCW)}, April
  2018, pp. 179--184.

\bibitem{7884970}
J.~Kua, G.~Armitage, and P.~Branch, ``A survey of rate adaptation techniques
  for dynamic adaptive streaming over http,'' \emph{IEEE Communications Surveys
  Tutorials}, vol.~19, no.~3, pp. 1842--1866, thirdquarter 2017.

\bibitem{8424813}
A.~Bentaleb, B.~Taani, A.~C. Begen, C.~Timmerer, and R.~Zimmermann, ``A survey
  on bitrate adaptation schemes for streaming media over http,'' \emph{IEEE
  Communications Surveys Tutorials}, pp. 1--1, 2018.

\end{thebibliography}

\end{document}